# LSTM networks for Music Generation

Xin Xu, Lanzhou University

**ABSTRACT**

The paper presents a method of the music generation based on LSTM (Long Short-Term Memory) and contrasts the effects of different network structures on the music generation.

## 1  Introduction

Music composition is considered creative, intuitive and therefore an inherent human ability. Nonetheless, it has a long history of mathematical approaches since Hiller and Isaacson proposed to use *Markov* chains for automatic composition [1]. Eck et al.[2] first lead LSTM to the music generation domain. They produced Blues generation system creating notes based on the front note sequences. Sturm et al.[3] first used Char-RNN using musical data of ABC format as training datasets. The current note vectors of One-Hot encoding could predict the probability distribution of the next note. They used it to generate the next character and finally created melody of a high quality. Lackner et al.[4] predicted the melody based on chords through training Char-RNN by separating chords and melodies. Chu et al.[5] came up with a hierarchical RNN, where the bottom layers generate the melody, while the higher levels produce the drums and chord. Choi et al.[6] presented the method using RNN to generate music with MIDI files.

The field of music generation includes a wide range of tasks such as the composition of melody, chord, rhythm[7], and even lyrics, i.e. every typical components of music, and has been subject to numerous research studies. Music can be represented as a sequence of events and thus it can be modelled as conditional probabilities between musical events. For example, in previous chord, while the whole chord progression often depend on the global key of the music. In many automatic composition systems, these relationships are simplified by assuming that the probability of the current state *p(n)* only depends on the probabilities of the states in the past *p(n−k) ... p(n−1)*. A sequence of musical events – notes, chord, rhythm patterns – is generated by predicting the following event given a seed sequence.

In this paper, I'd like to introduce applications of LSTM units for the music generation of single-track piano music with 70 MIDI files as datasets and other professors' research achievements.

## 2 The architecture

### 2.1 MIDI File

MIDI(Musical Instrument Digital Interface)[8] is a technical standard that describes a communications protocol, digital interface, and electrical connectors that connect a wide variety of electronic musical instruments, computers, and related audio devices for playing, editing and recording music. A single MIDI link through a MIDI cable can carry up to sixteen channels of information, each of which can be routed to a separate device or instrument. This could be sixteen different digital instruments, for example. MIDI carries event messages, data that specify the instructions for music, including a note's notation, pitch, velocity (which is heard typically as loudness or softness of volume), vibrato, panning to the right or left of stereo, and clock signals (which set tempo). And the small file size is one of its advantages.

Therefore, the notes in MIDI file could be set to correspond to a unique integer in the computer. In the paper, the project encode music sequences in this way with One-hot encoding.

### 2.2 RNN

Recurrent Neural Networks(RNNs), different from the feedforward neural networks, allow for incorporating long term dependency in the model. Theoretically, it can remember infinitely long sequences, although in practice it is limited by the vanishing gradient problem [9]. During the training of back-propagation through time, the gradient is extremely diminished by multiplications of sigmoid operations.

### 2.3 LSTM

LSTM (Long Short-Term Memory) units solved this vanishing gradient problem [9]. LSTM allows the gradient to be flowed by a separate path with not multiplication but addition operations. It can remember the context information for a relatively long time, adapted to data in time sequences.

Given a list of data in time sequences

$$X = \{X_t|_{t=1}^{T}\},$$

in every time step, there is an input vector $X_t \in X$. We suppose that these $t$ time steps are logical in some extents, like characters in language models, and need to learn there potential logical relation, the relationship of the context. Since every node in LSTM has a memory cell, for LSTM, the output vector $y_t$ at the $t$ time step depends not only the input $x_i$, but the input of previous $t$-$1$ time steps, $x_1$, …, $x_{i-1}$, whose effects on $y_t$ are decided by the implicit state $h_{t-1}$. Through the mechanism of information flowing controlled by input gates and forget gate, the memory cells in LSTM units need to remember or forget some information. The LSTM unit is as Figure 1.

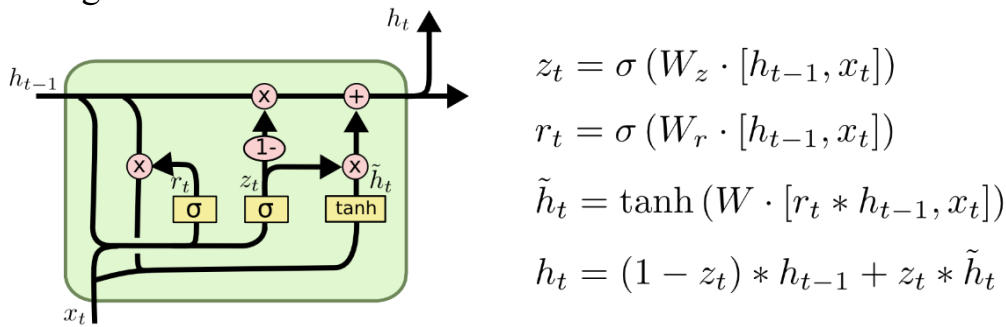

$$z_t = \sigma\left(W_z \cdot [h_{t-1}, x_t]\right)$$
$$r_t = \sigma\left(W_r \cdot [h_{t-1}, x_t]\right)$$
$$\tilde{h}_t = \tanh\left(W \cdot [r_t * h_{t-1}, x_t]\right)$$
$$h_t = (1 - z_t) * h_{t-1} + z_t * \tilde{h}_t$$

Figure 1: The LSTM unit[10]

## 3 The model of musical sequences with LSTM networks

### 3.1 The LSTM networks

The project uses three LSTM layers, two densely-connected layers. Each of LSTM layers consists of 512 hidden units and dropout of 0.3 is added after the first two LSTM layers and the first densely-connected layer. The first dense layer consists of 256 units and the number of units in the second dense layer is the same as the number of pitches from demos.

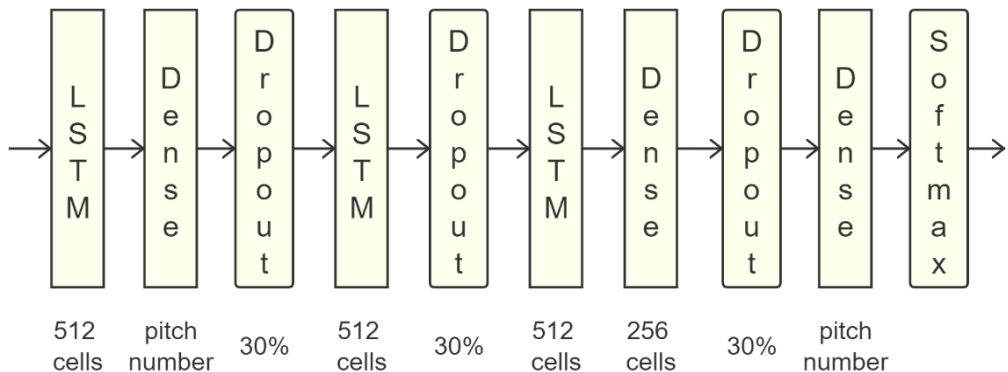

Figure 2 LSTM networks

The activation function the project chooses is *Softmax* function, the loss function is *Cross Entropy* and the optimizer is *RMSProp*. The model of the LSTM networks are as Figure 2.

### 3.2 The pre-processing of musical data

This project processes the single-track piano music. There are 70 piano demos, about 2 minutes for each.

Firstly, using *Music21*[11], the paper get every note and chord called *pitch* in these demos and organize them as a sequence stream. Next, the project splits them into some groups with 81 pitches per group. The front 80 pitches are set as a input sequence and the last one is the expected output pitch. Then the project creates a pitch-integer dictionary to store these pitches in the computer as integers. And the input is reshaped to a three-dimensional array and normalized to adapted to the *RMSProp* optimizer. Since the network use loss function *Cross Entropy*, the network output is need to be set as *One-Hot* encoding.

### 3.3 Train the network model

There is a function, *tf.keras.models.Sequential.fit()* in *Tensorflow*, which is used the network model in the project. The project sets the network input and out into the function as the parameter and run it. There is 100 epochs need to be accomplished. Meanwhile, there is also a checkpoint after every epoch recording parameters in the training process including the weights and the result of loss function. We can stop the training process anytime we want, like the time at which the loss is satisfied our expectation.

After training for 43 hours, 72 epochs, the loss rate is reduced to 0.4037 and the training is interrupted by person. The main reason is that the loss rate after 72 epochs increases because the model is not nice enough to control the learning rate leading to overfitting. Finally, after training, the project gets the best weights of the network to be used to generate music, which is from the epoch with the lowest result of loss function.

## 4  Music Generation

After training, the best weight of network can help to generate new music.

The training data is used in the network secondly to accomplish one forward propagation with the best weights and then we can get a predicted new music sequence, the output vector Y. The specific algorithm is as the following Generation Algorithm.

**Generation Algorithm**

Input: A random music sequence from training datasets and the trained
       LSTM network in Section 3.
Output: A predicted / generated music sequence file.
1) Load the best weight got in the training progress into the LSTM network.
2) Select a random music sequence $A$ as the input of the algorithm.
3) Predict the next music sequence based on the given sequence.
   The predicting algorithm is as the following:
       FOR i = 1, SEQLEN DO
           i.   Based on the input vector $A = [X_1, ..., X_{n-1}]$, generate the $n^{th}$ element $X_n$.
           ii.  Add the $X_n$ to the end of A.
           iii. Delete the first element in A.
                $A = [X_1, X_2, ..., X_n]$
       END
4) Process the generate music sequence and organize it into a musical piece.

## 5 Analysis and contrast of effects of different datasets, network model

### 5.1 The epoch time

Firstly, we look at the effect of the number of epochs on the loss. Figure 3 is the loss function in training process.

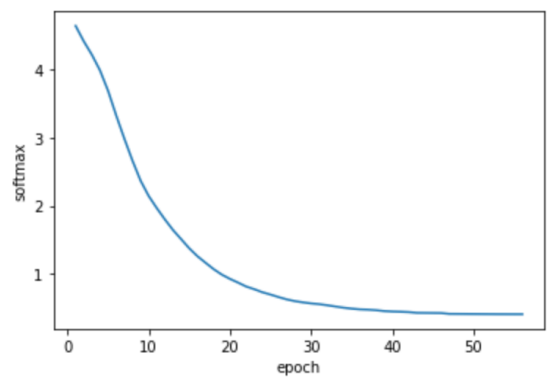

Figure 3 the loss

According to the result Figure 3, we can conclude that the deviation between the real output and the expected output gradually decreases as the epoch time increases. The more the epoch time is, the more the time of learning and adjusting weights is. Therefore, the epoch time can help improve the accuracy of the model in some extent. And the rate of decline in the loss function is slower after more epochs.

### 5.2 The number of neurons in the hidden layer of the network[12]

In the reference, the professor does 4-group comparing experiment and tests effects of numbers of neurons in the hidden layer of the network. The result is as Figure 4.

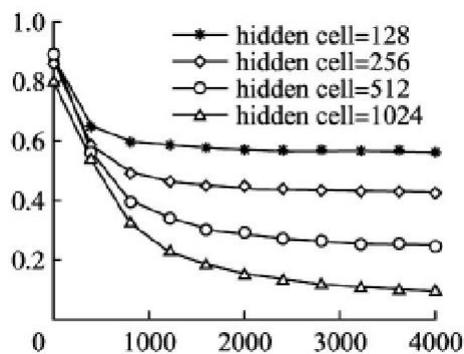

Figure 4: Effect of the number of neurons in the hidden layer of the network

The experiment proves that the more the number of neurons in the hidden layer of the network is, the stronger the ability of the LSTM network to learn the essence of datasets and abstract the characteristics is. And it can help reduce the deviation between the prediction and the expectation. The dimension of the hidden layer has a large effect on the quality of the generated music. However, deeper and wider networks need more calculation in the training process.